\documentclass{nic-series}

\begin{document}
\title{Homogeneous shear turbulence}
\author{Bruno Eckhardt, Andreas Dietrich, Arne Jachens and J\"org Schumacher}
\institute{Fachbereich Physik,
Philipps-Universit\"at Marburg, D-35032 Marburg, Germany}

\maketitle
\begin{abstracts}
The addition of suitable volume forces to the Navier-Stokes
equation allows to simulate flows in the presence of 
a homogeneous shear. Because of the explicit form
of the driving the flows are accessible to 
rigorous mathematical treatment and to accurate
quantitative modelling in their global properties. 
The statistics of the fluctuations provide insight
into generic behaviour of non-equilibrium systems
and into the presistence of anisotropies at small
scales. Correlation functions can be used to identify
dominant large scale dynamical processes that 
are relevant for most of the momentum transport
across the shear. The numerical studies of homogeneous
shear flows complement analytical and experimental 
investigations and contribute to bridging the gap between
ideal homogeneous isotropic turbulence and the more realistic
heterogeneous turbulence. 
\end{abstracts}

\section{Introduction}
The turbulent flow of a fluid is a paradigmatic example for
a non-equilibrium, nonlinear, and self-organizing system. The flow
has to be driven continuously to compensate the viscous 
friction due to shearing; the advection of the flow field
by itself introduces a characteristic nonlinearity;
and the flow shows a hierarchical organization of smaller
whirls on top of larger ones \cite{Fri95}. It is one of the aims
of turbulence theory to derive from the equations of motion 
the laws that govern this self-organization in space and time.
Partial progress has been made for 
the theoretically most appealing case of 
homogeneous, isotropic turbulence, where we have
the Karman-Horwath equation for third order moments \cite{Fri95}, 
a mean field theory \cite{GE},
and a variety of models\cite{Fri95,pope2000}, but for the most
part an analytical theory is not within reach. In the
absence of such a formalism numerical
simulations provide both access to the properties
of the equations as well as guidance in developing
intuition.

The power laws that describe the distribution of energy on the 
various scales quite well can be obtained
by dimensional arguments, following Kolmogorov, Onsager, v. Weizs\"acker,
Heisenberg and Richardson \cite{Fri95}.
These power laws are truncated on large scales by the 
forces that stir the fluid and on small scales by the
onset of viscous dissipation. Therefore, one would like to
study huge systems with infinitesimal viscosity in order
to obtain as large a scaling range as possible. In terms
of the large scale Reynolds number 
\begin{equation}
Re=UL/\nu
\end{equation}
formed by an external
velocity scale $U$, an external length scale $L$ and 
the viscosity of the fluid $\nu$, this is the limit of
$Re\rightarrow \infty$. Phenomenological arguments
show that the ratio between the externally imposed
length $L$ and the large eddy turnover time $T=L/U$ 
to the smallest scales $\eta$ and $\tau$
that appear in the dynamics
scale with Reynolds number like \cite{pope2000}
\begin{equation}
\eta/L\sim Re^{-3/4} \qquad \tau/T\sim Re^{-1/2}\,.
\end{equation}
Thus, the number of modes that are needed in order
to represent a 3-d flow field increases like $Re^{9/4}$.
Since the maximal time step in the integration is not
determined by the time scale of the flow, but 
by stability considerations following from the spatial
discretization, the number of time steps increases like
$Re^{3/4}$, so that the total operation count increases
like $Re^3$! As a consequence, the largest simulations run 
to date, on the Earth simulator with $4096^3$ modes, achieve
a Reynolds number of about 230.000 only, and follow the
flow field for just one large scale time unit \cite{japan}. 
However, the comparison of data by Sreenivasan \cite{Sreeni}
suggests that already with a resolution of about $256^3$ or $512^3$ 
one can reach into the inertial range behaviour and thus begin
to study the dominant statistical behaviour of turbulent 
flows. This then opens up the possibility to study flows
more realistic but also more complicated than homogeneous
isotropic turbulence.

In contrast to the theorists ideal, realistic flows 
are driven by anisotropic forcings
(e.g. a uni-directional winds), influenced by boundary
layers (near walls) and are in many other respects far from 
the case of homogeneous, isotropic turbulence.
A flow that is intermediate between realisitic flows
and the ideal cases, and that can be used to bridge the 
gap is turbulence with a homogeneous, linear
shear profile superimposed. It mimicks the situation
that one expects to find, at least locally. 
An investigation of such flows then allows to 
study the interaction between the turbulence and the
shear, the range over which the anisotropies affect
the scaling behaviour, and the consequences of the
shear for dynamical transport processes in the flow.

The problem with homogeneous shear flows is that it is
not obvious how to maintain the shear. Driving the fluid
from the boundaries, e.g. by oppositely moving side walls, 
will produce the
familiar boundary layers with large gradients 
and a bulk region with reduced ones. 
Rogallo \cite{rogallo1981} thus suggested to drive the fluid
by continuously shearing the computational grid. Obviously,
after a while, the grid will be extremely distorted and 
a remeshing will be necessary in order to restore the initial
resolution. When the appropriate terms are 
introduced into the Navier-Stokes equation, this amounts
to a periodic driving of the fluid, and thus a
temporally inhomogeneous situation.

In 1999 we 
started a research programme based
on a novel method with which we can simulate
homogeneous shear flows with prescribed forcing, shear,
and residual turbulence. The essential idea is to employ
a body force that can be tuned to fit the appropriate 
requirements. It is a numerical tool, not 
realizable in experiments, that is fairly versatile and
amenable to mathematical analysis and provides insights
impossible to obtain otherwise.

The outline of the article is as follows.
We begin in section 2 with a description
of the numerical methods used. We then turn in section 3
to a description of the global properties of the flow,
the connection to rigorous mathematical bounds and to
a model for the relaxation of the turbulent energy.
The fluctuations around the mean properties, their probability
distributions and their scaling with Reynolds number are
discussed in section 4. The dynamical properties as reflected
in dynamical correlation functions and their significance
for large scale momentum transport are discussed in section 5.
A brief outlook concludes the article.

\section{Simulating homogeneous shear flows}
The Navier-Stokes equation for an incompressible fluid 
driven by a divergence free volume force ${\bf F}$ is, 
in dimensionless form, 
\begin{eqnarray}
\label{nseq}
\frac{\partial{\bf u}}{\partial t}+({\bf u}\cdot{\bf \nabla}){\bf u}
&=&-{\bf \nabla} p+\frac{1}{Re}{\bf \nabla}^2{\bf u}+{\bf F}\;,\\
\label{ceq}
{\bf \nabla}\cdot{\bf u}&=&0\;,
\end{eqnarray}
with $p({\bf x},t)$ the pressure, and ${\bf u}({\bf x},t)$ the velocity field
that satisfies $\nabla\cdot{\bf u}=0$. The difficult and time-consuming
part of the integration is the evaluation of the nonlinear advection
term $({\bf u}\cdot{\bf \nabla}){\bf u}$. High Reynolds number simulations
therefore prefer Fourier representations for the velocity field, so that
Fast Fourier Algorithms can be used to obtain the gradients as a local
multiplication in wave number space. The aliasing problem is
handled with the 2/3 rule \cite{Orszag}.
In order to break the inherent
periodicity of the Fourier modes, which implies that to every
region with positive shear there will be one with negative shear 
shifted by half a period, we bound the domain by free-slip walls in the
shear direction: these boundary conditions are then compatible with 
a Fourier representation of the velocity field.

The coordinates are chosen such that $x$ points downstream, 
$y$ in the direction of the shear, and $z$ in spanwise direction. 
In $x$ and $z$ we take periodic boundary conditions, and in 
the $y$ direction free-slip conditions, i.e.,
\begin{equation} 
\label{bc}
u_y=0,\;\;\mbox{and}\;\;\frac{\partial u_x}{\partial y}=
\frac{\partial u_z}{\partial y}=0
\end{equation} 
at top and bottom surfaces. The size of the domain is 
$\lbrack 0,2\pi \rbrack \times \lbrack 0,L_y\rbrack\times
\lbrack 0,2\pi\rbrack$. The length $L_y$ is taken to be $1$ or $\pi$.
The typical Fourier resolution is $256\times 129\times 256$. As 
a rule of thumb the spectral resolution is considered sufficient
when the largest wave number $k_{max}$ and the smallest turbulent 
scales $\eta$ satisfy $k_{max}\eta >1$, although for quantities
sensitive to gradients the fluctuations in $\eta$ may have to 
be taken into account, so that even more restrictive conditions
have to be satisfied (Schumacher and Sreenivasan, work in progress).
Because of the dealiasing the largest wave number is given
by $\sqrt{2}N/3$ for $N$ modes \cite{Orszag}.

For turbulent flows 
it has become customary to use not the external Reynolds number
$Re$ but an easily computable and measurable
intrinsic Reynolds number $R_\lambda =U_{rms}\lambda/\nu$. 
It is based on the root mean 
square velocity $U_{rms}$ and the Taylor length scale $\lambda$ obtained 
from the gradients of the velocity field,
\begin{equation}
\lambda^{-2} = \frac{\langle \left( \partial u_x/\partial x\right)^2 \rangle}
{\langle u_x^2\rangle}\,.
\end{equation}
In isotropic turbulence the relation between the large scale 
Reynolds number $Re$ and the Taylor-Reynolds number $R_\lambda$ is given by
\cite{pope2000}
\begin{equation}
Re \approx 0.15 R_\lambda^2 \,.
\end{equation}
Our simulations reach up to $R_\lambda\approx 150$ on a
$512^3$ grid for the bounded flow domains, the 
earth simulator \cite{japan} 
manages $R_\lambda\approx 1200$ for isotropic turbulence.

The flow is naturally homogeneous in $x$ and $z$, so that for 
statistical purposes we can form averages over planes
$y=const$, henceforth denoted $\langle\ldots\rangle_A$.
It cannot be fully homogeneous in the $y$-direction because
of the vanishing gradients at top and bottom surfaces, but the
simulations show that the widths of these free slip boundary
layers is small and decreases with increasing Reynolds number
(see the inset of Fig.~\ref{fig2} and Ref.~7).

Two forms of driving are typically used: we can fix a force
field ${\bf F}$ and study the flow that results, or we can
prescribe the mean profile and adjust the force so that
this mean profile is maintained. In the latter case,
the force compensates the contributions that would 
result from the nonlinear term and the pressure, 
in order to keep the amplitudes of a set of
preselected Fourier components constant.

The mean profiles and the statistical properties of this
flow were compared to experiment and other simulations 
in \cite{schumachereckhardt2000}. It turns out that the 
method easily yields
a statistically stationary state and that the dynamics
of the flow is much less violent than in the case of
the Rogallo remeshing, i.e. the bursts in energy and 
enstrophy typically found there 
\cite{pumir1996,yakhot2003} are absent.

\section{Globally averaged properties}
A quantity that is amenable to rigorous mathematics without
supplementary hypothesis and additional models is the
total dissipation in the mean, i.e. the turbulent dissipation
averaged over the flow domain and over time,
\begin{equation}
\epsilon \equiv \nu \langle |{\bf \nabla u}|^2\rangle_{V,T} \,.
\end{equation} 
The optimal bound theories of Busse \cite{busse78}, the 
variational approaches of Constantin and Doering \cite{constantindoering}, 
Kerswell \cite{kerswell} and Nicodemus et al \cite{nicodemusetal} and 
their recent extension to volume driven flows 
\cite{doeringfoias,doeringetal2003} show that this
dissipation is rigorously bounded from above by an expression 
of the form
\begin{equation}
\epsilon
\le c_1\nu \frac{U_{rms}^2}{\ell^2}+c_2 \frac{U_{rms}^3}{\ell}
\label{upperb} 
\end{equation} 
with numerical coefficients $c_1$ and $c_2$. The velocity
scale $U_{rms}$ is set by the root mean square of the velocity and
$\ell$ is the length scale of the external driving. If a 
Reynolds number is formed with $U_{rms}$, i.e. $Re=U_{rms}\ell/\nu$,
then this bound is consistent with the expectation that
$\epsilon\sim Re^3$ for large $Re$.

\begin{figure}[b]
\begin{center}
\epsfxsize=7cm
\epsfbox{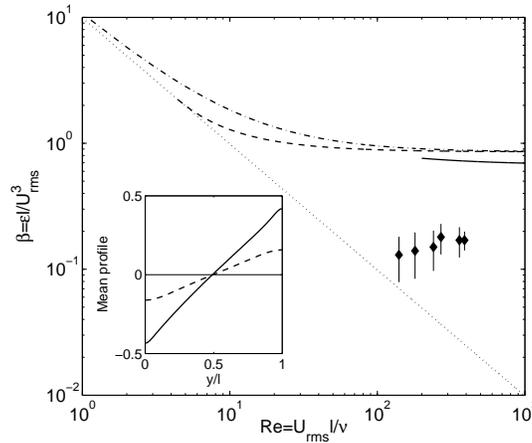}
\caption[]{Dimensionless dissipation ratio, $\beta$, vs. Reynolds number for a
flow driven by a constant shear force.
The results of the direct numerical simulations are indicated by diamonds 
with error bars from the statistics of the fluctuations. 
The straight dotted line indicates lower limit to the dissipation, 
obtained for a laminar shear flow. The three other lines are analytically 
derived and succesively improved bounds. The inset shows the mean flow profiles 
$\langle u_x\rangle_{A,T}$ vs $y$ for two different Reynolds numbers.
}
\label{fig2}
\end{center}
\end{figure}

For the specific case of a sinusoidal driving the bounds and the 
results from a numerical simulation are shown in Fig.~\ref{fig2}.
The coefficient from the upper bound is $c_2=\pi^2/\sqrt{216}\approx0.67$,
whereas the numerical simulations indicate $c_2\approx 0.2$, about
a factor 3 lower.
It is interesting to note that the difference between the bounds
and the numerical simulations is much smaller than in the case of 
wall driven shear flows \cite{nicodemusetal}, where the difference is
about a factor 10, and that in contrast to that 
case the dissipation does not tend to decrease with increasing
Reynolds number. This suggests
that volume forces provide a more efficient stirring than moving
boundaries.

\begin{figure}[b]
\begin{center}
\epsfxsize=12cm
\epsfbox{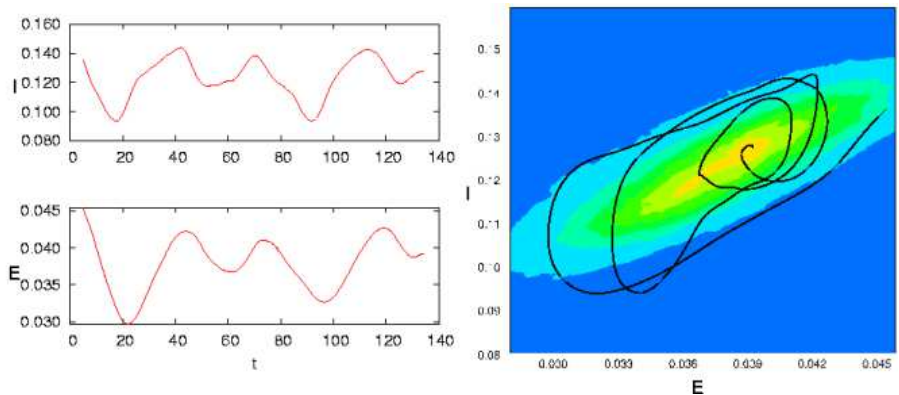}
\caption[]{Comparison of the energy model with direct numerical simulations of a shear flow. 
The left panels show time traces of the energy content (top) and energy uptake (bottom).
The panel on the right shows a typical trajectory from the DNS, superimposed
on the color contours of the probability density from the model in the presence of 
additive white noise.
}
\label{fig3}
\end{center}
\end{figure}

The instantaneous dissipation $\epsilon(t)$, 
energy content $E(t)=\langle{\bf u}^2\rangle_V/2$, and
energy uptake $I(t)=\langle{\bf u\cdot F}\rangle_V$
are not constant but fluctuate fairly irregularly.
An equation for their dynamics can be derived with less mathematical
rigor but still directly from the
Navier-Stokes equations with minimal assumptions. Multiply
the Navier-Stokes equation (\ref{nseq}) once with ${\bf u}$ and
once with ${\bf F}$ and average over the fluid volume. Then: 
\begin{eqnarray}
\frac{\mbox{d}E}{\mbox{d}t} &=& -\epsilon(t) + I(t)\,,\\
\frac{\mbox{d}I}{\mbox{d}t} &=& 
-\langle {\bf F}\cdot[({\bf u}\cdot\nabla){\bf u}]\rangle_V
+\nu \langle {\bf u} \cdot \Delta{\bf F}\rangle_V
+\langle{\bf F}^2\rangle_V\,.
\end{eqnarray}
In case the force field is an eigenfunction of the Laplacian, the
term $\langle {\bf u} \cdot \Delta{\bf F}\rangle_V$ becomes
proportional to $I(t)$.
The equations can then be closed by relating the energy dissipation and the
term quadratic in the velocity in the second equation to the energy content,
$\epsilon=c_d E^{3/2}$, and 
$\langle {\bf F}\cdot[({\bf u}\cdot\nabla){\bf u}]\rangle_V=c_f E$. Then 
\begin{eqnarray}
\frac{\mbox{d}E}{\mbox{d}t} &=& - c_d E^{3/2}(t) + I(t) \,,
\label{model1}
\\
\frac{\mbox{d}I}{\mbox{d}t} &=& - c_f E(t) - \nu\lambda I(t) + F \,,
\label{model2}
\end{eqnarray}
where the last term contains the norm of the force profile,
$F=\langle {\bf F}^2\rangle_V$. 
The model constants $c_d$ and $c_f$ can be determined from 
turbulent flow simulations. These equations have a stationary 
state, corresponding to the time average energy content
and energy uptake. Interestingly, the relaxation to this stationary
state is oscillatory. Fig.~\ref{fig3} shows a typical
trajectory from a numerical simulation:
the fluctuations that are superimposed on the 
mean values always drive the system away from the stationary
state, but the relaxation towards the stationary state
is not unlike the dynamics in the model. When the fluctuations
are modelled as additive white noise the colored 
probability density results. The intensity of the noise
was determined from the requirement that the second moments
of the measured fluctuations be reproduced.

The model provides a promising starting point for the analysis of
relaxational behaviour in excited turbulent flows (previously observed
in \cite{schmiegel,patches}) and for the response dynamics of
periodically driven flows \cite{bonn03,lohse03}. In view of the
exact results that could obtained for the stationary situation 
it will also be interesting to see whether similar results
can be obtained for periodically driven flows.

\begin{figure}[b]
\begin{center}
\epsfxsize=10cm
\epsfbox{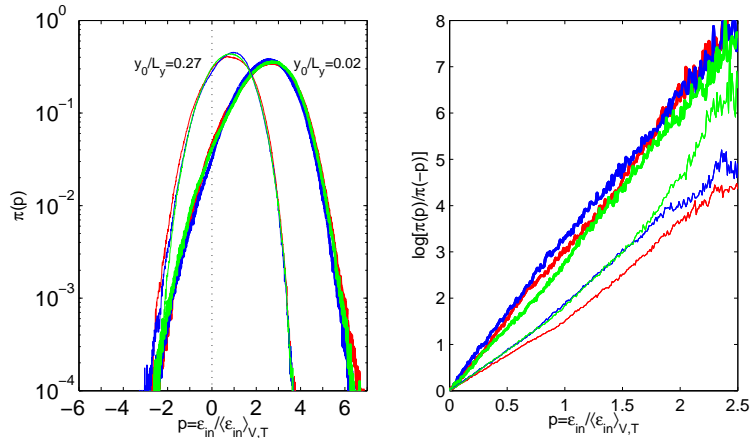}
\caption[]{Statistics of energy uptake for different vertical 
positions in the shear layer. The probability density functions (PDF)
normalized by the mean values are shown in the left panel.
They are based on more than 100 turbulent snapshots separated in time by
1.5 large eddy turnover times with a total of about $6\times 10^8$ data points.
The PDF's are shown for three different Reynolds numbers, $Re=270$ (red), 
$360$ (blue) and 
$390$ (green) and two different positions $y_0$ in the normal direction 
(thin lines for 0.27 and thick lines for 0.02).
The right panel shows the ratio of the probability for positive
and negative energy dissipations.}
\label{fig4}
\end{center}
\end{figure}

\section{Local fluctuations}
Simulations and model in the previous section show that 
energy dissipation and energy uptake in a turbulent flow
are fluctuating quantities, and even though they are equal
in the mean, they can be different for short times. 
The distribution of such fluctuations is a matter of debate,
with experiments by Pinton {\it et al.} \cite{Pinton98}
showing strongly non-Gaussian fluctuations, whereas
the data of Cadot {\it et al.} \cite{Cadot02} are
compatible with a Gaussian distribution. Since difference
may be related to the size of the domains over which the
flow field was averaged we studied the local energy uptake.
This also shows clear signatures for the presence of 
negative energy uptake, 
indicating a reverse flow in energy from the
turbulent fluid to the stirrer. Such events are very rare, 
and if the average over the fluid volume is taken, they
essentially never happen. However, locally 
the fluctuations of energy uptake rates can take on both
signs, and quite frequently become negative. 
Statistics of the instantaneous, 
pointwise energy injection rate 
$\epsilon_{in}({\bf x},t)={\bf u}({\bf x},t){\bf\cdot F}({\bf x})$ 
were analyzed in \cite{schumachereckhardt2003}. 
Since the system is invariant under translation in
downstream and spanwise direction, but not in the normal
direction, we study the distributions for planes 
parallel to the bounding free-slip surfaces separately.
The probability density functions of the energy input rate in units
of its ensemble average, $\langle\epsilon_{in}\rangle_{V,T}$, and for 
different positions between the plates are shown in Fig.~\ref{fig4}.  

The probability density functions for different Reynolds numbers collapse
nicely when normalized by the mean, but the distributions vary
considerably across the layer. Negative values, i.e. 
transfer of energy to the stirrer, occur fairly frequently, and
are more likely further away from the center. The relative frequency
between positive and negative energy uptake shows an exponential 
relation. Such exponential relations have recently been found in 
many non-equilibrium systems where they could be connected to 
fundamental symmetry properties of non-equilibrium invariant
measures \cite{evansetal93}. For hydrodynamic systems
these ideas do not strictly apply, since the Navier-Stokes equation 
is not reversible. 
But the study of Farago \cite{farago02} shows that even in the absence of that 
symmetry, as e.g. for a Brownian particle, similar relations
can be justified analytically.

Since homogeneous shear flows are in a sense the first step away
from ideal isotropic turbulence, one can ask how the statistical
properties at the smallest scales of the turbulent flow are affected by the
shear. In a pioneering paper Lumley \cite{lumley1967} predicted a 
rapid $R_\lambda^{-1}$ decay
of such anisotropies with Reynolds number. Recent systematic measurements in 
simple shear flows
\cite{gargwarhaft1998,ferchichitavoularis2000,shenwarhaft2000,staicuvandewater2003} 
for Taylor microscale Reynolds numbers up to $R_{\lambda}\sim 10^3$ show
clear deviations from the predicted decay. 
Moreover, direct numerical simulations (DNS) for moderate Reynolds
numbers confirm a persistent anisotropy and reveal
a relation to typical large-scale flow structures, 
so-called streamwise streaks
\cite{pumir1996,schumachereckhardt2000,schumacher2001,schumacheretal2003}. 
While the anisotropies still tend to become smaller with increasing
Reynolds number, the decay 
is slower than $R_\lambda^{-1}$.

\begin{figure}
\begin{center}
\epsfxsize=6cm
\epsfbox{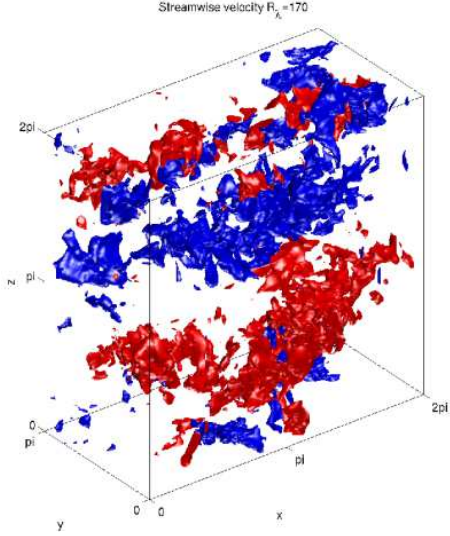}
\caption{Isosurface plot of a turbulent streamwise velocity snapshot for a DNS of
homogeneous shear turbulence at $R_{\lambda}=170$. Level at $v_x=2$ is coloured red,
level at $v_x= -2$ blue. The elongated structures can be identified as streaks.}
\label{fig5}
\end{center}
\end{figure}


\section{Dynamical correlations}
Various kinds of structures in shear flows such as
vortices, streaks or waves have been 
identified and considerable efforts have gone into identifying
their dynamical origins and evolution. The isosurfaces of the downstream
component of the flow field in Fig.~\ref{fig5} show that even
though the driving is homogeneous in spanwise direction 
the flow organizes into large regions with very high positive
(red) or negative (blue) downstream velocity. Such structures
are called streaks. They induce strong gradients
and shear instabilities and play an important role in a turbulent
recycling process described by Waleffe\cite{waleffe97} that 
consists of three steps:
i) downstream vortices 
mix fluid in the normal direction and drive modulations in the downstream
velocity, forming streaks; ii) streaks undergo an instability 
to the formation of vortices pointing in the normal direction; 
iii) the mean shear profile now turns these vortices again in 
downstream direction, thus closing
the loop. Of these processes the ones in step iii) and ii) are reasonably
fast, whereas the one in i) is fairly slow, since it is connected
with a non-normal amplification and thus only linear in time.

The indicator for non-normal amplification that we focus on here
is a temporal cross-correlation function \cite{eckhardtpandit03}. 
Since the vortex drives the streak a cross-correlation between the 
vortex and the streak should be asymmetric in time: if the streak 
is probed after the vortex then there might be a correlation, if 
it is probed before then there should not be a correlation. For 
the fluctuations in a linearization around a linear shear profile
this can be analyzed analytically \cite{eckhardtpandit03}.

\begin{figure}
\begin{center}
\epsfxsize=6cm
\epsfbox{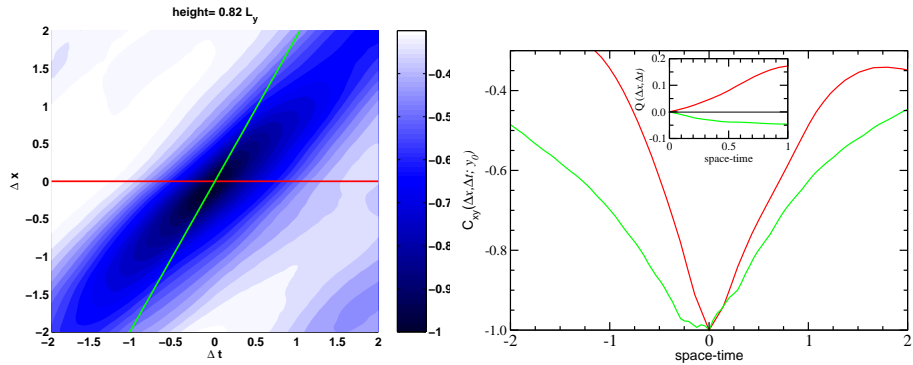}
\epsfxsize=6cm
\epsfbox{RadialCutAsym.eps}
\caption[]{Spatio-temporal cross-correlation function of a turbulent shear flow 
evaluated at a fixed normal position $y_0$ between the free-slip boundaries.
The left panel shows the contour plots of $C_{xy}(\Delta x, \Delta t; y_0)$ as
defined in (\ref{cross}). The right panel shows one-dimensional cuts
in time for a fixed position (red) and along a position moving with
the mean velocity, $\Delta x=\langle u_x\rangle(y_0)\Delta t$ (green). The
green correlation function is one to be expected on the basis of 
Taylor's frozen flow hypothesis \cite{townsend}.
The inset shows the asymmetry parameter $Q=(C(t)-C(-t))/(C(t)+C(-t))$.
}
\label{fig6}
\end{center}
\end{figure}

In order to see this in turbulent flows we study
Eulerian spatial and temporal cross-correlation functions
between the downstream and normal velocity components 
at fixed heights $y_0$, i.e.,
\begin{equation}
C_{xy}(\Delta x, \Delta t; y_0)=\langle
v_y(x, y_0, z, t) v_x(x+\Delta x, y_0, z, t+\Delta t)\rangle_{x,z,t} \,, 
\label{cross}
\end{equation}
The correlation function has elongated oval like isocontours that are
not aligned with the axes (Fig.~\ref{fig6}). The angle by
which it is tilted gives a velocity that seems to differ a bit from the 
mean velocity of the flow (indicated by the green line). 
One-dimensional cuts, such as the one along the red line
at fixed position, support an asymmetry, indicative of
the non-normal amplification process \cite{eckhardtetal03}.
The cut along the line given by the mean velocity (green)
is what would be expected in a hot-wire anemometor at a fixed
position using Taylors-hypothesis \cite{townsend}. 
 
\section{Outlook}
The investigations on homogeneous shear flows presented here have helped
to understand the dynamics of energy and dissipation, of the 
behaviour of the fluctuations and the dynamical process that are active.
There are several directions in which one can proceed:
one option is to add tracers in order to study the dynamics in 
advected Lagrangian frames, expecially the relative dynamics of 
three or more particles since it should contain information about
the statistically conserved quantities in the cascade \cite{Falkovich},
and since it connects to the experiments of Bodenschatz et al \cite{Bodenschatz}.
Along similar lines, it should be interesting to add scalar fields in order
to study their mixing dynamics \cite{SS_prl}. 
Active particles, like long
flexible polymers, will be stretched by the flow field \cite{cpc} 
and will
interact with the flow and will interfere with the large scale
dynamics, so as to reduce turbulent drag in the fluid. The insights
gained in the homogeneous shear flow studies will be extremely
valuable in understanding the properties of these added fields.

\section*{Acknowledgements}
We thank C.~R.~Doering, H.~H.~Fernholz, W.~I.~Goldburg, M.~Oberlack,
R.~Pandit, W. Schr\"oder, K.~R.~Sreenivasan, and P.~K.~Yeung for stimulating
discussions.   
Most of the simulations were done on the Cray T90 and Cray SV1ex 
supercomputers at the John von 
Neumann-Institut f\"ur Computing at the Forschungszentrum
J\"ulich and we are grateful for their steady support.  The work was also 
supported by the Deutsche Forschungsgemeinschaft,
by the European Network HPRN-CT-2000-00162 on Non-ideal turbulence, 
by the Alexander von Humboldt-Foundation within the Feodor Lynen Fellowship Program,
and by Yale University.

\end{document}